\documentclass[twocolumn,aps,floatfix,showpacs]{revtex4}
\usepackage{graphicx,epsfig}
\usepackage{dcolumn}
\usepackage{bm}
\usepackage{amscd}
\usepackage{amsmath}
\usepackage{amsfonts}
\usepackage{amssymb}
\usepackage{amsthm}
\usepackage{verbatim}
\usepackage{subfigure}

\newcommand{\be}{\begin{equation}}
\newcommand{\ee}{\end{equation}}
\newcommand{\ba}{\begin{array}}
\newcommand{\ea}{\end{array}}
\newcommand{\bea}{\begin{eqnarray}}
\newcommand{\eea}{\end{eqnarray}}
\newcommand{\bdm}{\begin{displaymath}}
\newcommand{\edm}{\end{displaymath}}

\theoremstyle{definition}

\begin{document}
\title{Dynamics of k-core percolation}

\author{C. L. Farrow, P. Shukla and P. M. Duxbury}
\email[]{duxbury@pa.msu.edu}
\affiliation{Physics and Astronomy Department, Michigan State University, East Lansing,
                            MI 48824, USA.}

\date{\today}
\begin{abstract}
In many network applications nodes are stable provided they 
have at least $k$ neighbors, and a network of k-stable nodes is 
called a k-core.  The vulnerability to random attack 
is characterized by the size of culling avalanches which 
occur after a randomly chosen k-core node is removed. 
Simulations of lattices in two, three and four dimensions, as 
well as small world networks, indicate that power-law
avalanches occur in first order k-core systems, 
while truncated avalanches are characteristic of second order cases.  
\end{abstract}


\maketitle

The k-core of a network is the set of nodes and edges which remain after 
all nodes which have less than $k$ neighbors, 
and the edges attached to them, have been culled 
or removed.  Percolation of k-cores has 
 applications ranging from magnetism (where it was invented and  
called bootstrap percolation)\cite{ChalupaJPC79,KogutJPC81}
and rigidity percolation\cite{Moukarzel1997,Duxbury1999}, 
to social networks\cite{Seidman1983} and protein networks\cite{Wuchty2005},
as well as the jamming transition\cite{Schwartz2006}. The behavior 
on Bethe lattices is particulary interesting due 
to the mixed nature of the transition where a first order 
jump in the order parameter is co-existent with a square 
root singularity\cite{ChalupaJPC79,Moukarzel1997}.
  This behavior has recently been 
associated with the jamming transition in 
granular media \cite{Schwartz2006} and has been analysed using a $1/d$ expansion, 
which confirmed that it persists in finite dimensions\cite{Harris2005}, and 
by large scale simulations, which suggest that in four dimensions the transition 
is not mixed\cite{Parisi2006}.
In general,  k-core 
percolation in finite dimensions is quite different than that 
occuring on random graphs, with 
metastability being a key new feature, so that for $k>z/2$ 
hypercubic lattices have thresholds approaching one in the 
large lattice limit\cite{AizenmanJPA88,ShonmannAP92,CerfSPA02,
HolroydPTRF03,DeGregorioPRL04}, whereas on Bethe lattices the k-core threshold 
occurs at values less than one for all $k\le z-1$\cite{ChalupaJPC79}.
Moreover on random graphs k-core percolation is first order 
for all $k>2$\cite{ChalupaJPC79} whereas on triangular lattices and cubic lattices 
the transition is second order and in the connectivity percolation 
university class for $k=3$\cite{ChavesPA95,MedeirosPA97,StaufferIJMPC96,
BrancoIJMPC99,AdlerBJP03,KurtsieferIJMPC03}. 

\begin{figure}
  \epsfig{file=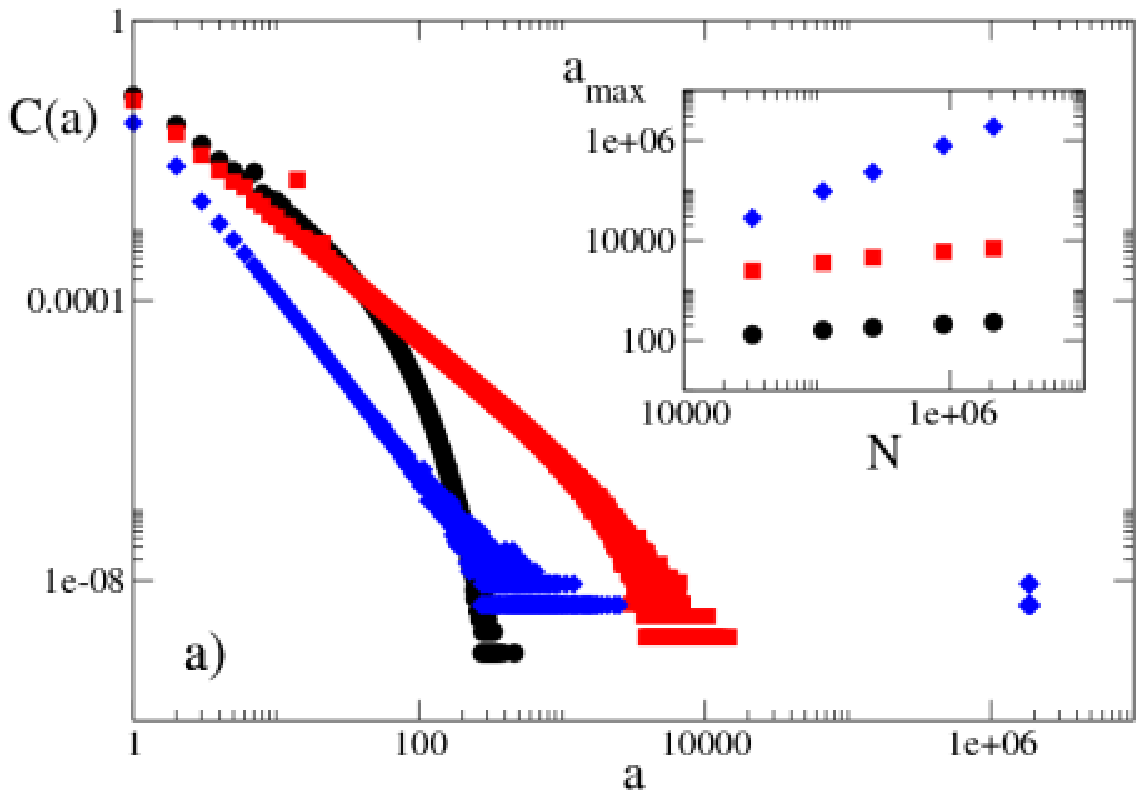 , width=0.65\columnwidth,angle=-90}
  \epsfig{file=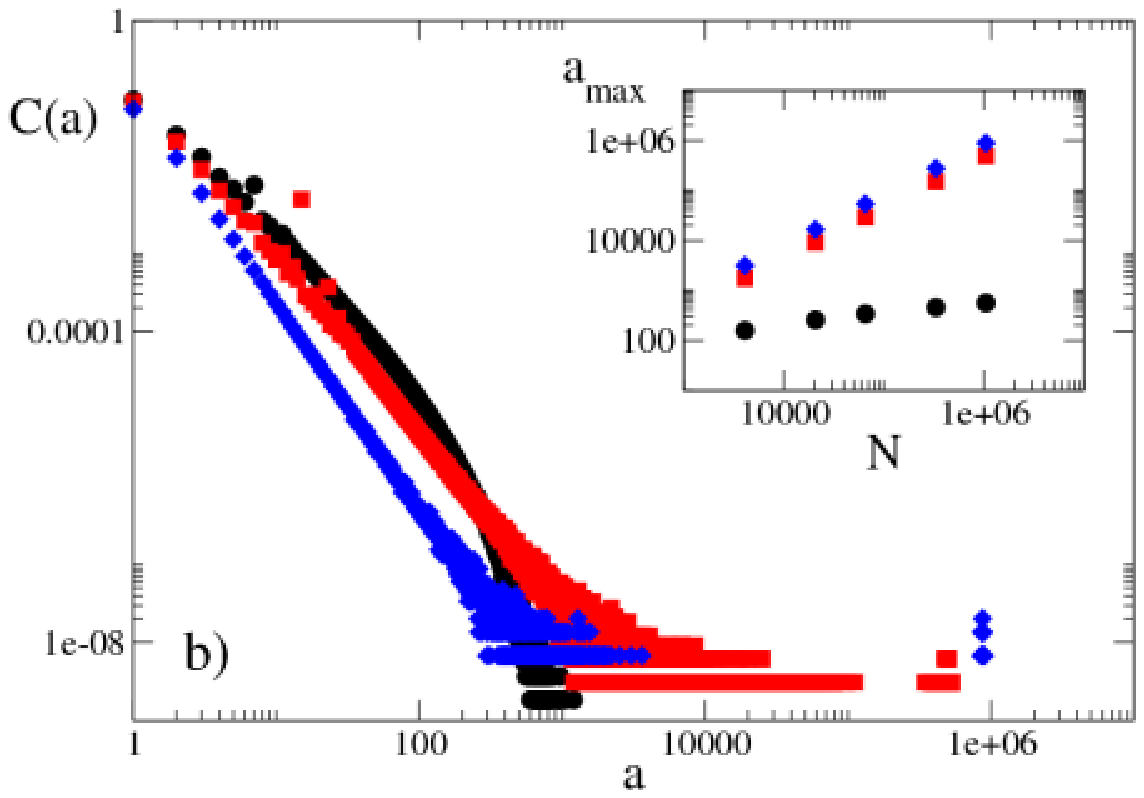 , width=0.65\columnwidth,angle=-90}
\caption{(Color online) Effect of lattice dimension (d),
 at fixed lattice co-ordination (8), on the avalanche distribution 
 (ECA), $C(a)$,  for the cases $k=3$(black circles),
 $k=4$(red squares) and $k=5$ (blue diamonds).
a) Body-centered cubic lattices ($d=3$) and, b) Simple cubic lattices ($d=4$). 
The $4$-core transition changes from second order for $d=3$
to first order for $d=4$ \cite{Parisi2006}, as is clearly seen 
in the avalanche distributions and in the finite 
size scaling of the largest avalanche, $a_{max}$ (Insets).  
In a) $1000$ samples were used, and in the main figure the lattice size $L=128$. 
In b) $1000$ samples were used, and in the main figure $L=32$.}
\end{figure}
The vulnerability of networks to random node removal or attack, 
has a variety physical applications, including: radiation damage; 
species extinctions in biological networks;
and random outages or cascading failures in
communication or transportation networks\cite{albert2000}.
There are also relations to avalanches in materials problems,  
including fracture and earthquakes\cite{Sethna2001,alava2006}, and Barkhausen noise 
in magnetic materials\cite{Dahmen2001,Sethna2001,Durin2000}. In fact bootstrap percolation 
is related to the weak disorder limit of 
one of the key models of disordered magnets, 
the random field Ising model (RFIM)\cite{SabhapanditPRL02}. 
\begin{figure}
\epsfig{file = 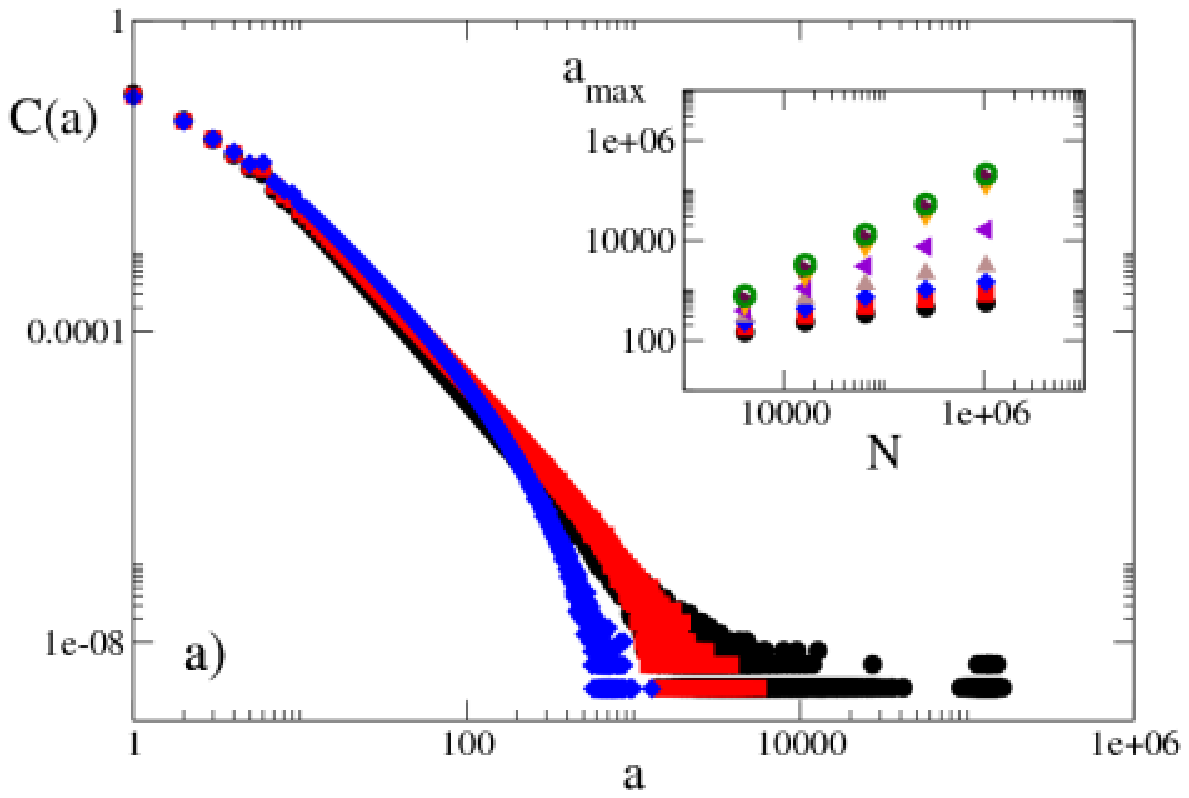,
				width=0.65\columnwidth,angle=-90}
\epsfig{file = 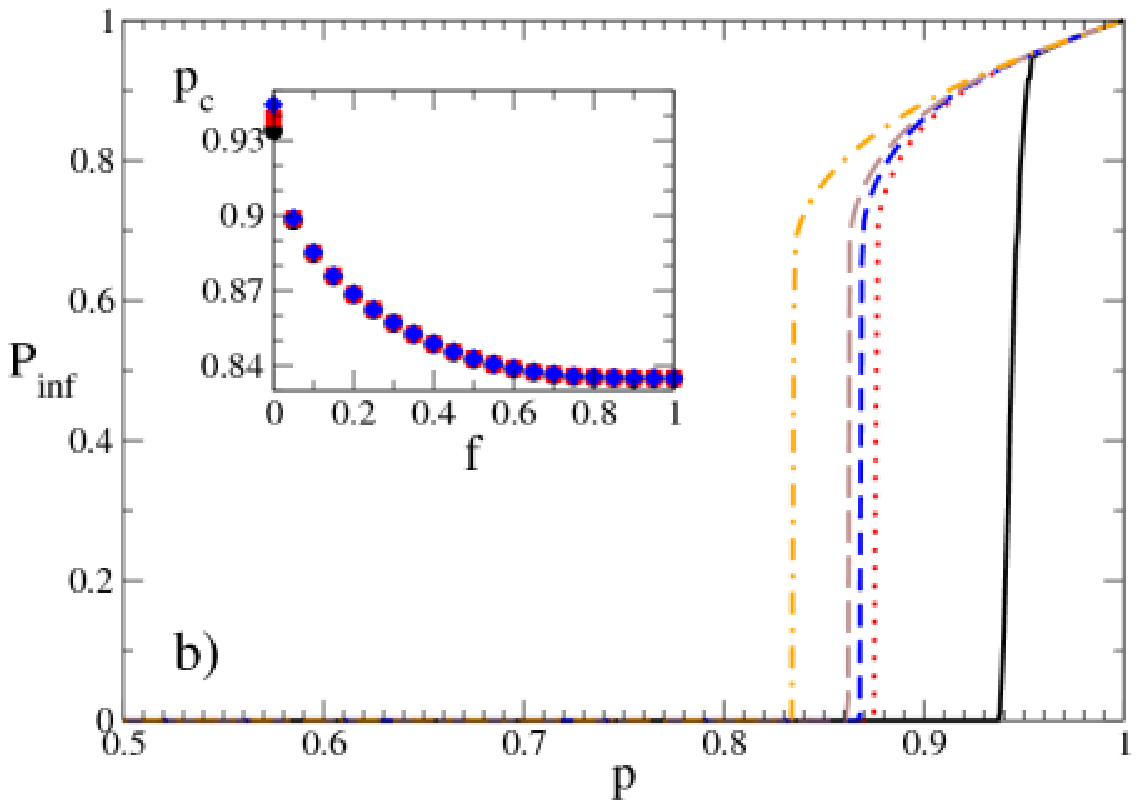,
				width=0.65\columnwidth,angle=-90}
\caption{(Color online) k-core percolation on six co-ordinated small world 
networks, with fraction $f$ of small world connections.
 a) Avalanche distributions, $C(a)$, for 3-core cases  
with $f=0.15$ (blue diamonds), $f=0.20$ (red squares) and $f=0.25$ (black circles).  Inset: 
The largest avalanche, $a_{max}$, starting at the bottom, for $f=0.15,0.17,...,0.25$. 
b) The infinite 4-core probability, $P_{inf}$ versus site probabiltiy $p$ for small world fractions (from right to left) $f=0, 0.15, 0.20, 0.25,1$.  
Inset: the bootstrap percolation threshold, $p_c(f)$.
There is a singular behavior in $p_c(f)$ as $f\rightarrow 0$ (see text). 
In the main figures of a) and b) and in the inset to a) 
from $1000$ samples for each $f$, with $L=1028$.  In the inset to b) from 100 samples 
for $L=256$(black circles), $L=512$ (red squares), $L=1024$(blue diamonds).}
\end{figure}

In an earlier paper\cite{FarrowPRE2005}, we defined elementary culling avalanches 
(ECA) in k-core or bootstrap problems to be avalanches induced by random removal 
of one node on a stable k-core, which is a dynamical procedure analogous to 
that used in crackling noise simulations in random field systems\cite{Sethna2001}.  
We showed that on triangular and cubic lattices the ECA distribution 
provides a sensitive indicator of the nature of the bootstrap transition, 
even in the slowly convergent second order case $k=3$ \cite{ChavesPA95,MedeirosPA97,StaufferIJMPC96,BrancoIJMPC99,AdlerBJP03,KurtsieferIJMPC03}
and the metastable $k=4$ case\cite{AizenmanJPA88,ShonmannAP92,CerfSPA02,HolroydPTRF03,DeGregorioPRL04}.

Here we show that in systems ranging from 
small world networks to regular lattices, including hypercubic lattices 
in four dimensions, the dynamics of k-core percolation 
falls into two classes: (i) Second order systems which have truncated 
avalanche distributions where the largest avalanche grows very 
weakly with system size and; (ii) First order systems where 
avalanches are power law distributed and the largest avalanche 
is of order the system size, producing a first order jump in the 
infinite cluster probability at percolation.
  In the case of Bethe lattices, 
 power law avalanches in first order cases
can be demonstrated explicitly\cite{Dorogovtsev2006} (see below).

Culling cascades for a variety of lattices 
in finite dimensions are presented in Figs. 1 and 2a. 
These distributions are found by starting with a 
lattice with all sites present and then repeating   
 two steps:  (i) Randomly remove a site and all edges 
connected to it, (ii) Recursively cull all unstable sites until a stable k-core 
is achieved.  The number 
of sites removed during step (ii) constitutes an elementary culling avalanche(ECA).
The distribution of such avalanches, beginning with an undiluted lattice 
and continuing until the lattice is empty yields the cummulative culling avalanche 
distribution, $C(a)$.  We also monitor the 
largest avalanche that occurs for a given lattice, $a_{max}$, which is the 
maximum cascade over the whole avalanche trace.  The finite size scaling 
behavior of the maximum cascade is presented in the insets to Figs. 1 and 2a.

In Fig. 1 we present avalanche distributions  
for $k=3,4,5$, on eight-coordinated lattices in three (Fig. 1a)
and four dimensions (Fig. 1b).  These 
lattices where chosen to demonstrate the effect of spatial dimension on the nature 
of the k-core transition.  It is clear from this data that the $k=4$ transition 
changes from second order to first order as we go from three to four dimensions
as noted recently\cite{Parisi2006},   
however the $k=3$ transition is second order in both three and four dimensions. 
In contrast on $z=8$ Bethe lattices and 
random graphs, k-core percolation is first order for all $k>2$. 
Clearly dimension plays an important role on the nature of the 
k-core transition, moreover whenever the transition is first order 
we see the same characteristic behavior: power law avalanches and 
a largest avalanche of order the sample size\cite{FarrowPRE2005}.  

We investigated a second method for changing the nature 
of k-core transitions, namely by changing 
the range of the interactions.  To do this, we 
created small world networks by starting
with an undiluted triangular lattice and randomly removing 
a fraction $f$ of bonds from it. 
These bonds are then randomly put back into the lattice, 
but they may connect any pair of sites at arbitrary distances, with the constraint 
that all sites remain six-coordinated. In this way we create six co-ordinated 
small world networks with fraction $f$ of long range connections.
 When $f$ is zero, we have a regular 
triangular lattice, while when $f=1$, we generate a random regular graph 
with co-ordination six everywhere.  By varying $f$, we monitor the effect 
of long-range connections on the nature of the k-core 
transition, which also measures the 
 vulnerability of the small world networks 
to random attack\cite{albert2000}.  For $k=3$, 
we find that, as illustrated in Fig. 2a, the transition changes 
from second order to first order when the fraction of small world connections 
reaches $f_c = 0.23(2)$. For $f>f_c$ there are power law avalanches and 
an extensive largest avalanche, while for $f<f_c$, the avalanche 
distribution is truncated, with no large avalanches.  

The 4-core transition on 6-coordinated networks is 
first order for all $f$, however on triangular lattices 
it is metastable so that for large lattices the 
threshold approaches one very slowly, while on random graphs 
there is a finite threshold $p_c <1$. Nevertheless we find 
power law avalanches and an extensive largest avalanche in 
all of these cases.  However, the behavior of the k-core threshold as 
a function of the fraction of small world connections is interesting 
and is presented in Fig. 2b.  A notable feature of this 
data is that small world connections quickly stabilize the metastable 
4-core of triangular 
lattices and, as is emphasized in the inset to Fig. 2b, there is 
a singular behavior in $p_c(f)$ as $f\rightarrow 0$.  Presumably, this non-perturbative 
behavior is related to the stabilization of large holes in the 
triangular lattice by the addition of long-range loops characteristic of small world
networks, and it may be possible to characterize this behavior using 
rigorous mathematical methods \cite{AizenmanJPA88,ShonmannAP92,CerfSPA02,HolroydPTRF03}.

We also observed power law avalanches and 
an extensive largest avalanche in the limit $f=1$ 
which is the random graph limit.  In this limit 
power law avalanches can be demonstrated explicitly 
using Bethe lattices, as we now show. 
Consider site diluted, z-coodinated Bethe lattices where a 
fraction $p$ of the nodes are randomly removed, and 
let $T$ be the branch probability that a node  
of the Bethe lattice is part of a k-core. The branch probability, $T$, 
is found by solving the self-consistent equation,
\begin{equation}
T = p\sum_{n=k}^{z-1} A^{z-1}_nT^n; \ \ with\ \ A^r_l = {r\choose l} (1-T)^{r-l}
\end{equation}
which is well known from the original bootstrap paper\cite{ChalupaJPC79} and 
from applications to rigidity percolation\cite{Moukarzel1997}.
The branch probabilities are used to find the probability 
of a site being on the k-core, 
\begin{equation}
X = p\sum_{n=k}^{z} A^z_nT^n.  
\end{equation}
We build on the results (1) and (2) to find the culling 
avalanche distributions for a branch, $B(a)$, 
and for the Bethe lattice, $P(a)$.  $P(a)$ is 
the differential probability and is related to 
the cummulative probability, $C(a)$ through
 $C(a) = \int P(a,p) dp$.
\begin{figure}[h!] 
\includegraphics[width=0.7\columnwidth]{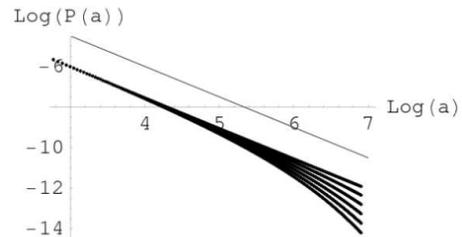} 
\caption{A log-log plot of culling avalanches, $P(a)$ 
for 3-core percolation on a z=4 Bethe lattice, found from $P^*(y)$
(see text after Eq. (8) 
using an exact Mathematica algorithm.  The dark lines are 
for 5 values of the site concentration, from the top starting with 
$p=8/9=p_c$ and then moving above $p_c$ in increments of $\delta p = 10^{-4}$.
The continuous line of slope 3/2, vertically shifted from the data, is included to demonstrate 
that $P(a)\sim 1/a^{3/2}$ at the 3-core threshold, $p_c$.}
\end{figure}

As noted previously\cite{Dorogovtsev2006,Goltsev2006}, sites which have exactly 
$k$ neighbors play a special role in culling avalanches on Bethe lattices.  Following 
their terminology, we define the corona as a connected cluster of 
sites each of which is part of the k-core and has exactly k neighbors.
With this definition, removal of a site of the k-core  
leads to a culling avalanche in the corona surrounding that site. 
The probability of finding a corona of size, $a$, surrounding 
a randomly chosen starting site on the Bethe lattice is given by,
\begin{equation}
P(a) = p\sum_{n=k}^z A^z_n 
\prod_{l=1}^n \sum_{a_l} B(a_l)\delta(S_n).
\end{equation}
where $S_n = \sum_{l=1}^n a_l - a+1$.
This expression has the same form as Eq. (2), but with the term $T^n$ in that 
equation replaced by the product in Eq. (3).  This term is a product over the 
probabilities of finding corona on each of the branches connected to the 
starting site, and the delta function ensures that the ECA's on 
these corona sum to the total avalanche size $a$.  From this equation 
it is evident that the normalizations $\sum P(a) = X$ and $\sum B(a) = T$ hold.
A similar reasoning is applied to the branch probabilities, yielding 
the extension of Eq. (1) to, 
\begin{equation}
{B(a)\over p} = \sum_{n=k}^{z-1} A^{z-1}_nT^n \delta(a)  
+  A^{z-1}_{k-1} \prod_{l=1}^{k-1} \sum_{a_l} B(a_l)\delta(S_{k-1}).
\end{equation}
There are two terms in this expression, as only corona sites  
on a branch initiate a culling avalanche on the branch.  All other k-core 
sites terminate the ECA and contribute to the prefactor of $\delta(a)$ in this equation.
This delta function is not present in Eq. (3) as the starting site there is removed 
by hand to initiate an ECA so it can have any co-ordination. 
Defining the generating functions, 
$B(y) = \sum B(a) y^a$, and $ P(y) = \sum P(a)y^a$, 
and applying them to Eq. (3) yields, 
\begin{equation}
P(y) = p y \sum_{n=k}^z A^z_n B(y)^n.
\end{equation}
Similarly, Eq. (4) becomes, 
\begin{equation}
B(y) = p\sum_{n=k}^{z-1}  A^{z-1}_n T^n  
+ p y A^{z-1}_{k-1} B(y)^{k-1}.
\end{equation}
The probabilities $B(a)$ or $P(a)$ are extracted from the 
generating functions $B(y)$ or $P(y)$, by differentiation or 
by contour integration, whichever is more convenient.  

The cases $k=1,2$ are the standard connectivity 
percolation and percolative backbone problems respectively, 
so the simplest non-trivial k-core problem on random graphs
is $k=3$, $z=4$.  The formalism outlined above can be carried through explicitly 
in this case and has a behavior characteristic of all cases $k\ge 3$
\cite{Dorogovtsev2006,Goltsev2006}. 
For this case, $B(y)$ obeys, 
\begin{equation}
B(y) = p T^3 + 3py(1-T)B(y)^2, 
\end{equation}
where $T = 3/4 + 9(\delta p)^{1\over 2}/8 \sqrt 2$, which follows from Eq. (1) 
and $\delta p = p-p_c$.  This illustrates the characteristic feature of a 
jump discontinuity and a continuous square root singularity, as noted 
before\cite{ChalupaJPC79,Moukarzel1997}.
Solving Eq. (7) and selecting the physical root, we find an 
explicit expression for the avalanche generating function,
\begin{equation}
B^*(y) = {1-[1-12p^2T^3(1-T) y]^{1\over 2} \over 6p(1-T) y}.
\end{equation}  
and finally, from Eq. (4), we find the generating 
function for Bethe lattices, 
$P^*(y) = p y [4(1-T)B^*(y)^3 + B^*(y)^4]$.
We wrote a Mathematica program which finds the exact 
culling avalanche distribution from $P^*(y)$, and the results are 
presented in Fig. 3, which demonstrates the $3/2$ power law behavior 
at threshold characteristic of mean field avalanches.
Asymptotic analysis near criticality 
is carried out by noting 
that the normalization condition $B^*(y=1) = T$, enables 
an expansion of $B^*$ near $y=1$ and $p=p_c$.
The leading term in the expansion of $B^*$ is then,
$B^* \approx  1-[1-y(1+9\delta p/2)]^{1/2}$,
from which we find that near the threshold, $p_c$, the 
tail of the avalanche distribution obeys
$P(a) \sim [1-D(p-p_c)]^{a}/a^{3/2}$,
where  $D = 9/2$ and $p_c=8/9$ for 3-core percolation 
on 4-coordinated Bethe lattices. 
In experiments as well as computer simulations, it is more convenient to measure
the integrated probability distribution of avalanches $C(a)$ and 
in the simulation results presented above, we integrate over the 
complete avalanche trace. The probability of large avalanches is
significant in  a small region of width $\frac{1}{\delta p}$ near threshold.
Integration of $P(a)$ in this regime yields, 
$C(a) \sim a^{-\frac{5}{2}}$, for $ a \rightarrow \infty$ 
which is observed in our simulations on random graphs, and 
is typical of avalanches occuring in infinite range models 
and Bethe lattices in a wide variety of non-equilibrium systems 
\cite{Kloster1997,Vespignani1998,Sethna2001,alava2006,Goltsev2006}. 

In all cases we studied, power law avalanches are
associated with first order k-core transitions, both in 
metastable and regular first order cases.  
This is not without precident as in random field Ising 
hysteresis a first order jump occurs in the hysteresis loop 
despite the fact that power law avalanches occur \cite{Dahmen2001,Sethna2001}.  
However in brittle fracture problems in finite dimensions, 
 the avalanches are truncated 
when the loading sharing is short ranged\cite{Hidalgo2002,Zapperi2005}.  Presumably this 
is due to the greater amplification of stress on survivors in fracture 
problems with short range load sharing.  Nevertheless,  
k-core dynamics capture the essential physics of instabilities 
in a broad class of complex network problems.
 

\end{document}